\documentclass[11pt, a4paper]{article}

\usepackage{mathtools}
\usepackage{amsfonts}
\usepackage{amssymb}
\usepackage{graphicx, rotating}
\usepackage{epstopdf}
\usepackage{epsfig}
\usepackage{latexsym}
\usepackage{color}
\usepackage{cite}
\usepackage{slashed}
\usepackage[hyperfootnotes=true]{hyperref}
\definecolor{bluscuro}{rgb}{0.15, 0.2, .85}
\definecolor{ndc}{rgb}{0.7, 0.3, .3}
\hypersetup{colorlinks, citecolor=black, linkcolor=black, urlcolor=bluscuro}
\usepackage[font={small,it}]{caption}
\usepackage{geometry}
\usepackage{subcaption}

\definecolor{GRed}{rgb}{0.9,0.18,0.2}
\definecolor{CBlue}{rgb}{0.1,0.12,0.9}

\definecolor{amber}{rgb}{1.0, 0.49, 0.0}
\definecolor{auburn}{rgb}{0.43, 0.21, 0.1}

\newcommand{\nn}{\nonumber}

\newcommand{\be}{\begin{equation}}
\newcommand{\ee}{\end{equation}}
\newcommand{\bea}{\begin{eqnarray}}
\newcommand{\eea}{\end{eqnarray}}

\newcommand{\gsim}{\lower.7ex\hbox{$\;\stackrel{\textstyle>}{\sim}\;$}}
\newcommand{\lsim}{\lower.7ex\hbox{$\;\stackrel{\textstyle<}{\sim}\;$}}

\numberwithin{equation}{section}
 
\def\pplogo{\vbox{\kern-\headheight\kern -29pt
\halign{##&##\hfil\cr&{\ppnumber}\cr\rule{0pt}{2.5ex}&\ppdate\cr}}}
\makeatletter
\def\ps@firstpage{\ps@empty \def\@oddhead{\hss\pplogo}%
  \let\@evenhead\@oddhead 
}
\def\maketitle{\par
 \begingroup
 \def\thefootnote{\fnsymbol{footnote}}
 \def\@makefnmark{\hbox{$^{\@thefnmark}$\hss}}
 \if@twocolumn
 \twocolumn[\@maketitle]
 \else \newpage
 \global\@topnum\z@ \@maketitle \fi\thispagestyle{firstpage}\@thanks
 \endgroup
 \setcounter{footnote}{0}
 \let\maketitle\relax
 \let\@maketitle\relax
 \gdef\@thanks{}\gdef\@author{}\gdef\@title{}\let\thanks\relax}
\makeatother

\begin{document}

\setcounter{page}0
\def\ppnumber{\vbox{\baselineskip14pt
IPPP-15-62, DCPT-15-124}
}
\def\ppdate{\footnotesize{}} \date{}

\title{\bf Convexity, gauge-dependence and tunneling  rates
}
\thispagestyle{plain}
\maketitle

\begin{center}
\vspace{0.7cm}
{ \large  Alexis D. Plascencia and}  {\large  Carlos Tamarit}\\[7mm]
{ \it  Institute for Particle Physics Phenomenology}\\
{\it Durham University, South Road, DH1 3LE, United Kingdom}\\

\vspace{1.cm}
{\it email: a.d.plascencia-contreras@durham.ac.uk, carlos.tamarit@durham.ac.uk}\\
\vspace{1.cm}
\today
\end{center}
\vspace{1.5cm}
\begin{abstract}

We clarify issues of convexity, gauge-dependence and radiative corrections in relation to tunneling rates.
Despite the gauge dependence of the effective action at zero and finite temperature, it is shown that tunneling and nucleation rates remain independent of the choice of gauge-fixing. Taking as a starting point the functional that defines the transition amplitude from a false vacuum onto itself, it is shown that decay rates are exactly determined by a non-convex, false vacuum effective action evaluated at an extremum. The latter can be viewed as a generalized bounce configuration, and gauge-independence follows from the appropriate Nielsen  identities. This holds for any election of gauge-fixing that leads to an invertible Faddeev-Popov matrix.  
\end{abstract}

\newpage


\section{Introduction}

Since the work of Jackiw and Dolan \cite{Jackiw:1974cv,Dolan:1974gu}, it has been known that the quantum effective action in gauge theories, and in particular its zero momentum piece, the effective potential, 
depend on the choice of
gauge-fixing. The effective
potential is used to calculate physically meaningful quantities, both at zero temperature (such as vacuum energies, masses, tunneling rates), as well as at finite temperature
(e.g. critical temperatures in phase transitions and their nucleation rates). Given that physical observables cannot depend on the choice of gauge, it becomes important to understand how to extract 
gauge-independent information from the effective action. 

The works of Nielsen, Kugo and Fukuda \cite{Nielsen:1975fs,Fukuda:1975di} set the basis for the resolution of these issues, providing identities that encode the behavior of the effective action under changes of the gauge-fixing parameter. Originally
derived for specific classes of gauge-fixing functions, these identities have been extended to arbitrary choices of the latter \cite{Voronov:1982ph,Kobes:1990dc} \footnote{See also \cite{Lavrov:1981kb,Lavrov:1981ag,Aitchison:1983ns,Johnston:1984sc,
DoNascimento:1987mn,Contreras:1996nx,DelCima:1999gg,Alexander:2008hd} for discussions about the validity of the Nielsen identities in a variety of gauges.}. The Nielsen identities imply that the  gauge dependence of 
the effective action is equivalent to a nonlocal field redefinition. For the effective potential, in the case of a single scalar field $\sigma$, they adopt the form 
\begin{align}
\label{eq:NielsenV}
 \left(\xi\frac{\partial}{\partial\xi}+C(\sigma;\xi)\frac{\partial}{\partial \sigma}\right)V_{\text{eff}}(\sigma;\xi)=0,
\end{align}
where $\xi$ is the gauge-fixing parameter, $V_{\text{eff}}$ the quantum effective potential, and $C(\sigma;\xi)$ is a functional which can be calculated in terms of Feynman diagrams. An immediate consequence of this is that physical 
quantities defined at extrema of the effective potential, where $\partial V_{\text{eff}}(\sigma;\xi)/\partial\sigma=0$, become 
gauge-independent \cite{Nielsen:1975fs,Aitchison:1983ns}. This is the case for vacuum energies, as well as masses of scalar fluctuations around vacua. However, vacuum expectation values of fields, as well
as the values of the effective potential in between minima, remain gauge-dependent and hence unphysical. The Nielsen 
identities have 
been explicitly studied and verified for several theories, both at zero and finite temperature, mostly for scalar QED in a variety of gauges (see e.g. \cite{Aitchison:1983ns,Johnston:1984sc,Johnston:1986cp,DoNascimento:1987mn,
Metaxas:1995ab,
Contreras:1996nx,Metaxas:2000cw,Alexander:2008hd,Binosi:2005yk,Garny:2012cg,Andreassen:2014eha}), but also in the Standard Model \cite{Andreassen:2014gha}.  Being a nonperturbative result, some care has to be taken to define a 
perturbative counting scheme such that 
the identities hold order-by-order. This can happen for example with the vacuum energy in a truncated perturbative calculation. Given that the minimization conditions may enforce a relation between powers of the tree-level 
couplings of the theory that differs from the usual loop counting (such as in Coleman-Weinberg models of radiative symmetry breaking \cite{Coleman:1973jx}), some resummation of loop effects might be needed to explicitly 
check the gauge independence of the vacuum energy to a given level of approximation \cite{Patel:2011th,Andreassen:2014eha,Andreassen:2014gha}.

The issue of the gauge dependence of vacuum energies and physical masses being solved by the Nielsen identities, one may worry that the gauge parameter might still work its way into calculations that depend on the values of the potential
away from the vacua, such as tunneling or nucleation rates. These are needed to tackle important questions in  particle physics, such as  the stability of the Higgs vacuum during and after 
inflation (see e.g. \cite{DiLuzio:2014bua,Bednyakov:2015sca,Lalak:2015usa,Espinosa:2015qea} for references including discussions on gauge dependence), or the properties of phase transitions in the early Universe,
which can have
an impact on baryogenesis (see \cite{Morrissey:2012db} for a review).  Some important studies of the gauge dependence
of tunneling amplitudes have been done in previous works, using semiclassical techniques and focusing either 
in the action of the corresponding Euclidean solution \cite{Metaxas:1995ab,Metaxas:2000cw}, or in the determinants of the fluctuations around the latter \cite{Baacke:1999sc}. In these works, gauge independence 
was shown to hold at the lowest nontrivial orders in 
perturbation theory for specific theories and choices of gauge-fixing, yet these analyses could not discard the appearance of gauge dependence at higher orders. Reference \cite{Garny:2012cg} found
a nonzero gauge dependence of tunneling rates in the Abelian Higgs model, but this was interpreted as a possible effect of the breaking of the derivative expansion of the effective action.
As a possible solution to the issue of the gauge dependence of the effective action, which would seem to allow to compute gauge-independent tunneling rates,  Nielsen has advocated \cite{Nielsen:2014spa} for the use of a potential obtained by 
performing a field redefinition that compensates the gauge dependence of \eqref{eq:NielsenV}. A related simplified approach, in which the scalar fields in the effective action are canonically normalized  by absorbing field-renormalization factors, was used in \cite{Lalak:2015usa,Espinosa:2015qea,Rose:2015lna}. Still, it remains unclear to see how this redefined potential
could arise  in the calculation of tunneling  rates from first principles.

The problem of the gauge dependence of tunneling rates is connected to that of including quantum corrections in a consistent manner in the calculation of tunneling amplitudes. In the usual computations
by means of a saddle point expansion of the path integral, the role of the effective potential --which itself includes quantum corrections--
is in principle unclear. This becomes especially problematic in theories with vacua that only arise radiatively, with the classical potential appearing inside the path integral having 
no nontrivial extrema. From early on it was assumed that the right answer involved using the effective potential, rather than its classical counterpart \cite{Frampton:1976kf}, though it was not until
the work of \cite{Bender:1984jc,Weinberg:1992ds} that the correctness of this procedure was justified in part. There it was argued that one could compute tunneling rates by doing the usual semiclassical expansion in an effective theory obtained by integrating out the gauge fields. However, it was also noted in \cite{Weinberg:1992ds} that the resulting effective potential does not exactly match the full
effective potential of the theory, among other things because it is obtained from connected, rather than one-particle-irreducible Green functions, and it does not contain fluctuations of the scalar fields. Thus the exact role played by the effective potential, as well as  the consequences of its  gauge dependence, remained unclear. Very recently, a formalism for consistently calculating tunneling rates by performing a saddle point evaluation of the path integral around the quantum, rather than classical path, was developed and applied in \cite{Garbrecht:2015cla,Garbrecht:2015oea,Garbrecht:2015yza}. Again, a clear justification for the use of the quantum path, and an understanding of the ensuing consequences for the gauge dependence of the results when considering theories with gauge fields, is still missing.

A further puzzle related with the possible role of the effective action in the computation of tunneling rates is related to its known reality and convexity properties \cite{RevModPhys.47.165}. The convexity (negative second derivative everywhere) of the effective action implies concavity (positive second derivative everywhere) of the effective potential, which thus cannot have false vacua. Furthermore, the true effective potential lacks an imaginary part, which would be associated with  an unstable state. This suggests that  another quantum functional must play a role in the calculation of tunneling rates, such as one of the ``localized'' effective actions proposed by Weinberg and Wu \cite{PhysRevD.36.2474}.\footnote{``Localization'' refers here to a requirement of a small dispersion in the expectation values of field operators.} It should be noted that the assumption that the true-vacuum functional does not play a role in tunneling rates is implicitly made whenever the rates are computed for unstable potentials without a true vacuum. Reference \cite{PhysRevD.36.2474} constructed a localized effective potential by 
constraining the size of quantum fluctuations, and interpreted its imaginary part as encoding the decay probability for the localized states. However, the implementation of the constraints on quantum fluctuations has ambiguities, and a higher-order definition is lacking. Well-defined constrained vacuum functionals are used in lattice theory, as the constraint effective potential used in \cite{O'Raifeartaigh:1986hi}, and introduced in \cite{Fukuda:1974ey}. However, the constraint effective potential reduces to the usual effective potential in the infinite volume limit, and so once more we don't expect it to play a role in tunneling rates. 

To the best of our knowledge, a nonperturbative result concerning the gauge independence of tunneling rates and their relation to an action functional is lacking. In this paper  we remedy this by providing a simple derivation based on a generalization of Callan and Coleman's  definition of the tunneling rate \cite{Callan:1977pt}.  We start by introducing false vacuum functionals associated with the transition amplitude from a false vacuum onto itself,  from which one can derive false vacuum effective actions which can be understood as a generalization of Weinberg and Wu's localized functionals. The latter effective actions remain complex and non-convex\footnote{Throughout this paper, we use the following conventions, matching those in \cite{RevModPhys.47.165}. ``Convex'': negative second derivative everywhere. ``Concave'': Positive second derivative everywhere. ``Non-convex'': Neither concave nor convex.}, and their gauge-dependence is encoded by their associated Nielsen identities. The tunneling rate is then expressed in terms of the imaginary part of the Euclidean false vacuum effective action evaluated at a generalized bounce solution to its quantum equations
of motion. This validates the approach of \cite{Garbrecht:2015cla,Garbrecht:2015oea,Garbrecht:2015yza}, which is shown to enforce the correct boundary conditions in the vacuum-to-vacuum path integral. Gauge-independence follows from the Nielsen identities of the false vacuum effective action, which imply that its value on extremal configurations does not depend on the gauge parameters.

Much like S-matrix elements are independent of the choice of gauge, and may be calculated with an arbitrary choice of gauge-fixing, tunneling rates can be computed in any gauge. Nielsen's field redefinition for arbitrary gauges can be thought of as a transformation of the fields which takes them to a reference gauge slice, and is not essential for achieving gauge-independent results. The cancellation of the gauge dependence is automatic --up to higher order
effects in a perturbative truncation-- as long as the effective action is evaluated consistently, including derivative terms. 

The organization of this paper is as follows. First we introduce vacuum functionals in section \ref{sec:vacfun}. Section \ref{sec:effac} is devoted to effective actions, and section \ref{sec:Nielsen} to their Nielsen identities.  Finally, tunneling rates are considered in section \ref{sec:tunneling}. Conclusions are drawn in \ref{sec:conclusions}.

\section{\label{sec:vacfun}Vacuum functionals}

We will start by studying the properties of vacuum functionals defined in terms of path integrals. These functionals are the vacuum-to-vacuum transition amplitudes in the presence of sources --both for the true vacuum and for a false vacuum-- and  their associated effective actions, obtained by means of  Legendre transformations.  We will emphasize that decay rates are associated with the false vacuum functionals, rather than the ones corresponding to the true vacuum of the theory. This explains why one can consistently consider false vacua and their  decay rates, including radiative corrections, despite the reality and convexity of the true vacuum effective action, which prevent it from playing a role in the calculation of decay rates. 


We will consider a theory with fields labelled in DeWitt's compact notation \cite{DeWitt:1967ub} as 
$\phi\equiv\{\phi_j\}$, with the index $j$ referring to any continuous or discrete degree of freedom, including space-time dependence. In the presence of a gauge symmetry with a Lie
Algebra ${\mathfrak g}$ spanned by generators $T^a,a=1\cdots{\rm dim}({\mathfrak g})$, there will be gauge transformations under which the classical action will be invariant. These transformations depend on a gauge 
parameter $\alpha=\alpha^a T^a$, and can be written as
\footnote{To ease the notation we assume a simple gauge group with a positive definite metric acting on the Lie Algebra, and work in the basis in which it is given by the identity. Hence we do not need to distinguish 
between upper and lower indices in the Lie Algebra, though we keep the distinction in the field indices in DeWitt's notation.}
\begin{equation}
\label{eq:gaugetr}
 \delta\phi_j\equiv D^a_j[\phi] \alpha^a.
\end{equation}
Let's first consider the vacuum-to-vacuum transition amplitude in the presence of a source, $Z[J]$. We will assume that the source produces the same perturbation, yielding the same groundstate, at times $t=\pm\infty$. The functional $Z[J]$ is related by the generator $W[J]$ of connected amplitudes  as $Z[J]=\exp i W[J]$. Introducing a complete basis of Heisenberg-picture, time-independent eigenstates $|q\rangle$ of the field operators $\hat\phi$, such that $\hat\phi_i|q\rangle=q_i|q\rangle$, the identity operator can be written as
\begin{align}
 I=\int [dq]\mu(q)|q\rangle\langle q|,
\end{align}
where $\mu$ is an integration measure. Using the above spectral decomposition, we may write the vacuum-to-vacuum amplitude as
\begin{equation}
\label{eq:W0}
\begin{aligned}
 Z[J]=\exp i W[J]=\lim_{T\rightarrow\infty}\langle 0| e^{-iHT}|0\rangle^J=\lim_{T\rightarrow\infty}\int [dq] [dq']\mu(q)\mu(q')\langle0|q\rangle^J\langle q|e^{-iHT}|q'\rangle^J\langle q'|0\rangle^J\\
 =\int [dq][ dq']\mu(q)\mu(q')\psi^J_0(q')\psi^{J\star}_0(q)\int_{q'}^q[d\phi]\mu(\phi)\exp{i\left[\tilde S_g[\phi;\xi]+J_j\phi^j\right]}\equiv\langle\exp[i J_j\phi^j]\rangle.
 \end{aligned}
\end{equation}
In the above equation, we remind the reader that the trace over $j$ indices includes an integration over space-time coordinates. $\psi^J_0(q)=\langle q|0\rangle^J$ can be understood as a field-space wave-function of the vacuum state in the Heisenberg picture and in the presence of the source $J$. The  integration measure $\mu$ is required to satisfy \cite{Kobes:1990dc}
\begin{align}
\label{eq:mu}
 \mu_{,j} D^{aj}+\mu D^{a j}_{,j}=0.
\end{align}
This happens for example in dimensional regularization (DR) with a constant $\mu$, since the $D^{aj}$ are linear in the fields, and then $D^{aj}_{,j}$ becomes an integral of a constant function which vanishes in DR. $\tilde S_g[\phi;\xi]$ in equation \eqref{eq:W0} is given by the classical action plus a gauge-fixing piece, depending on a gauge-parameter $\xi$, on which we will elaborate later. Finally, the notation for the integration symbol in $\phi$ in \eqref{eq:W0} alludes to the fact that the fields must satisfy the following boundary conditions,
\begin{equation}
  \lim_{t\rightarrow-\infty} \phi=q',\quad   \lim_{t\rightarrow\infty} \phi=q.
\end{equation}

Since the vacuum-to-vacuum transition amplitude is a phase, analytic continuation to Euclidean time $T_E=i T$ allows to define real Euclidean functionals $Z_E[J]=\exp W_E[J]$, where $Z_E[J]$ can be identified with the average of a positive function with a real measure. Indeed, doing the analytic continuation of equation \eqref{eq:W0}, one has
\begin{equation}
 Z_E[J]=\langle \exp J^i\phi_i\rangle.
\end{equation}
From this, using H\"older's inequality applied to positive functions with a real measure, reference \cite{RevModPhys.47.165}  argued that the generator of connected diagrams $W_E[J]=\log Z_E[J]$ is a real, concave functional, i.e. satisfying $(1-\alpha)W_E[J_1]+\alpha W_E[J_2]\geq W_E[(1-\alpha)J_1+\alpha J_2]$ for $0\leq\alpha\leq1$. Continuing back to Minkowski space-time, this implies that $W[J]$ is concave as well.

Typically, it is assumed that the wave-functional of the vacuum peaks around a single point in field-space, $\psi^J_0(q)\sim \delta(q-q_0^J)$, so that $Z[J]$ can be expressed as a single path integral
\begin{equation}
\label{eq:Wapprox}
 Z[J]\approx \int_{q_0^J}^{q_0^J}[d\phi]\mu(\phi)\exp{i\left[\tilde S_g[\phi;\xi]+J_j\phi^j\right]}.
\end{equation}
However, this  approximation will fail in the presence of nearly degenerate $N$ multiple vacua, in which case one expects the true vacuum's wave-function to peak around the field configurations $q^{J,m}_0,m=1,\dots, N$ of the local vacua. Since the energies of the vacua depend on the external current, near degeneracy will always be attained for some value of the current. For example, the classical potential in the presence of a current is modified to $V(\phi)-J_i\phi^i$, so that for different values of $J$ different vacua will be preferred. For these reasons $Z[J]$ will be better approximated by a sum of path integrals, as in
\begin{align}
\label{eq:Wsum}
 Z[J]\approx\sum_{m, n=1}^N Z^{m,n}[J],\quad Z^{m,n}[J]= {\cal N}[J]_{mn} \int_{q^{J,m}_0}^{q^{J,n}_0}[d\phi]\mu(\phi)\exp{i\left[\tilde S_g[\phi;\xi]+J_j\phi^j\right]},
\end{align}
where the ${\cal N}[J]_{mn}$ are current-dependent normalization constants related to the size of the peaks on top of the different vacua in the vacuum wave-function.  For values of the current for which there is a clearly preferred vacuum, one expects a single peak, and so one will recover the usual single path integral formula. However, this won't be true for all values of $J$, and for different values of $J$ the single-integral limits may come from different path integrals. Explicit calculations in the literature show that 
a single path integral  fails to yield a concave $W[J]$, and yet the sum over path integrals --interpreted in general as a sum over saddle-points-- gives a concave $W[J]$ whenever it is a good approximation to the full $Z[J]$ \cite{Fujimoto:1982tc,Bender:1983nc,Cooper:1983cd,Tabata:1985hu,Hindmarsh:1985nc,Alexandre:2012hn,Alexandre:2012ht}.\footnote{As noted in some of these references, this is similar to the Maxwell construction in Thermodynamics,  which gives a concave free energy as a result of the coexistence of phases; in quantum field theory one gets a concave effective potential as a result of quantum superposition.} In the former works it was also shown how in different regions in $J$ for which one of the vacua is clearly preferred, one recovers  single path-integral limits, as argued before. Here we reinterpret the sum over saddle-points as a sum over peaks of the vacuum's wave function. 

In the presence of false vacua, apart from $Z[J]$ one may introduce an analogous functional corresponding to the transition of an unstable state (or false vacuum) onto itself. This functional will play a role in the definition of the tunneling rate from the false vacuum. Denoting this unstable state by $|F\rangle$ and its field-space wave-function by $\psi^J_F$, then we may write, in analogy with equation \eqref{eq:W0},
\begin{equation}
\label{eq:WF}
\begin{aligned}
 Z^T_F[J]=\langle F| e^{-iHT}|F\rangle^J=
 \int [dq][ dq']\mu(q)\mu(q')\psi^J_F(q')\psi^{J\star}_F(q)\int_{q'}^{q}[d\phi]\mu(\phi)\exp{i\left[\tilde S_g[\phi;\xi]+J_j\phi^j\right]}.
 \end{aligned}
\end{equation}
In the previous formula, for finite values of $T$ the time integrals implicit in the last exponential are assumed to be taken for $-T/2\leq t\leq T/2$. Since the state is unstable and decays, the Hamiltonian acting on it picks an imaginary part, and one cannot obtain a real functional by analytic continuation to imaginary time. Thus, in contrast to the true-vacuum case, $Z^T_F[J]$ cannot be related with an average of a positive real functional, and one cannot use the arguments of \cite{RevModPhys.47.165} to prove concavity of $W^T_F[J]=-i\log Z^T_F[J]$. 

The complex functional $Z^T_F[J]$ allows to calculate the decay rate of the false vacuum. The false vacuum will be an approximate eigenstate of the Hamiltonian, with the corresponding energy eigenvalue picking an imaginary part \cite{Callan:1977pt}. Then, considering a normalization such that the energy of the false vacuum state is zero, $Z^T_F[0]$ in equation \eqref{eq:WF} can be written as
\begin{align}
\label{eq:WF0}
Z^T_F[0]=\langle F| e^{-iHT}|F\rangle\sim e^{-i\epsilon VT},
\end{align}
where $\epsilon$ denotes the false vacuum energy density. An instability is signalled by an imaginary part of $\epsilon$, which yields an associated decay rate
\begin{align}
\label{eq:tauG}
 \gamma=-2\,{\rm Im}\epsilon=-\lim_{V,T\rightarrow\infty}\frac{2}{VT}\,{\rm Re}\,(\log Z^T_F[0]).
\end{align}
Note that an unstable vacuum is associated with an imaginary $W^T_F[0]=-i \log Z^T_F[0]$, in contrast to the true-vacuum functional $W[0]$ which remains real.
Again,
whenever the false vacuum's wave-function peaks at a field configuration $q_F$, one may approximate $Z^T_F[J]$ by a single path integral,
\begin{equation}
\label{eq:WFapprox}
 Z^T_F[J]\approx \int_{q^J_F}^{q^J_F}[d\phi]\mu(\phi)\exp{i\left[\tilde S_g[\phi;\xi]+J_i\phi^i\right]}.
\end{equation}
As was commented in regards to equation \eqref{eq:Wsum}, a single path integral will fail to yield a concave functional $W^T_F[J]$, as is expected for the false vacuum.

Before moving on to the construction of effective actions, some comments are in order. Our definition of the tunneling rate from the false vacuum functional is slightly different from Callan and Coleman's
\cite{Callan:1977pt}. 
These authors start with the transition amplitude $\langle q',t'| q,t\rangle$  between generic eigenstates of the field operators. When inserting the identity operators expressed as a sum over 
projectors into the energy eigenstates, it is argued that in the $T\rightarrow\infty$ limit the transition amplitude is dominated by the exponential with minimum energy, which they associate with the false vacuum. This procedure is sometimes questioned, as the $T\rightarrow\infty$ limit could pick up the true-vacuum state rather than the false vacuum. Here we avoid the problem by starting with the transition
amplitude of the false vacuum onto itself, rather than a generic state $|q\rangle$. The false vacuum is an approximate eigenstate of the Hamiltonian, and thus its overlap with the true vacuum is suppressed, (going to zero as the false vacuum becomes long lived) and one cannot argue that in the infinite time limit one is left only with the contribution from the energy of the true-vacuum. In the single-path integral approximation, one is effectively considering an amplitude of the form $\langle q_F,t'|q_F,t \rangle$, but now $|q_F,t \rangle$ is not a generic state, but rather one with a maximum overlap with the false-vacuum. Furthermore, the true-vacuum cannnot contribute a real  part to $\log Z^T_F$, and so by restricting to ${\rm Re}\log Z^T_F$ and then taking the $T\rightarrow\infty$ limit, it is ensured that only the false-vacuum can contribute. Note also that although $Z^T_F$ goes to zero at $T\rightarrow\infty$, because the false-vacuum decays, its logarithm does not, and thus \eqref{eq:tauG} is a sensible definiton.

\section{\label{sec:effac}Effective action functionals}

From the above vacuum functionals, one may construct effective action functionals that depend on the mean fields by performing Legendre transformations. The usual mean field $\bar\phi_j\equiv \langle \phi_j\rangle^J$ represents the expectation value of the field $\phi_j$ in the groundstate and in the presence of a source, and is defined as
\begin{align}
\label{eq:Wav}
\bar\phi_i=\langle \phi_i\rangle^J=\frac{\delta W[J]}{\delta J^i}=e^{-iW}\sum_{m,n=1}^N {\cal N}[J]_{mn}\int_{q^{J,m}_0}^{q^{J,n}_0}\,[d\phi]\mu(\phi)\phi_i
 \exp i\left( \tilde S_g+J^j\phi_j\right),
\end{align}
where we have used the approximation of equation \eqref{eq:Wsum}, in which the vacuum functional is given by a sum of  path integrals with boundary conditions determined by the $N$ peaks of the vacuum wave-function. One may also define a false vacuum mean-field $\bar\phi^T_F$, which, using the approximation \eqref{eq:WFapprox},  will be given by
\begin{align}
\label{eq:WFav}
\bar\phi^T_{F i}=\langle \phi_i\rangle_{F}^J=\frac{\delta W^T_F[J]}{\delta J^i}=e^{-iW^T_F}\int_{q^J_F}^{q^J_F}\,[d\phi]\mu(\phi)\phi_i
 \exp i\left( \tilde S_g+J^j\phi_j\right).
\end{align}
The effective action $\Gamma$ is given by
\begin{align}
\label{eq:GammaW}
 \Gamma[\bar\phi]=W[J]-J_j\,\bar\phi^j ,
\end{align}
where it is understood that the mean fields and the sources are related by the following identities,
\begin{align}
\label{eq:W}
 \bar\phi_j=\frac{\delta W[J]}{\delta J^j},\quad\quad J_j=-\frac{\delta \Gamma[\bar\phi]}{\delta\bar\phi^j}\equiv-\Gamma_{,j}.
\end{align}
If the vacuum functional can be written as in equation \eqref{eq:Wsum}, it can be seen that $\Gamma$ may be implicitly defined by the following sum of path integrals (where we generalize the single path-integral results of references \cite{'tHooft:1975vy,Boulware:1980av,Abbott:1980hw} to account for a multi-peaked vacuum wave-function)
\begin{equation}
\label{eq:Gamma}
 \exp i\Gamma[\bar\phi;\xi]=\sum_{m,n=1}^N{\cal N}_{mn}[\Gamma_{,j}]\int_{q^{J,m}_0-\bar\phi_\infty}^{q^{J,n}_0-\bar\phi_\infty} [d\phi] \,\mu(\phi)\exp i\left[S_g[\bar\phi,\phi;\xi]-\Gamma_{,j}[\bar\phi;\xi]\phi^j\right].
\end{equation}
Note how the boundary conditions in the sum of path integrals depend on the combinations $q^{J,m}_0-\bar\phi_\infty$, where $\bar\phi_\infty$ represent the limiting values of $\bar\phi$ at $t\rightarrow\pm\infty$ (we are considering sources which lead to the same groundstate at those times). This arises after performing field-redefinitions inside the path integrals that define $Z[J]$, which is reflected by the change in notation in $S_g$, such that $S_g[\phi,\bar\phi;\xi]=\tilde S_g[\bar\phi+\phi;\xi]$, with $\tilde S_g$ appearing in  equations \eqref{eq:W0}, \eqref{eq:Wsum}, \eqref{eq:WF}, \eqref{eq:WFapprox}, \eqref{eq:Wav}, \eqref{eq:WFav}. Given the relation between $J$ and $\bar\phi$ enforced by equation \eqref{eq:W}, the $q^{J,m}_0$ can be expressed as a function of the mean field $\bar\phi$.
Using the definition \eqref{eq:GammaW} of the effective action,  it can also be seen that \eqref{eq:Wav} can be rewritten in terms of $\Gamma$ as
\begin{align}
\label{eq:av}
 \bar\phi_i=\langle\phi_i\rangle^J= e^{-i\Gamma}\sum_{m,n=1}^N{\cal N}_{mn}[\Gamma_{,j}]\int_{q^{J,m}_0}^{q^{J,n}_0}\,[d\phi]\mu(\phi)\phi_i
 \exp i\left[ \tilde S_g[\phi;\xi]-\Gamma_{,j}(\phi^j-\bar\phi^j)\right].
\end{align}
Given the concavity of $W[J]$, $\delta W/\delta J$ has a monotonous dependence on $J$ and is thus a single-valued functional. This implies that $\tilde\Gamma[J,\bar\varphi]\equiv W[J]-J\bar\varphi$, when considered as a function of $J$ for a fixed $\bar\varphi$ (with $J$ and $\bar\varphi$ unrelated), is concave and has a unique minimum at $J$ satisfying $\delta W/\delta J=\bar\varphi$, so that 
\begin{equation}
 \label{eq:Gammamin}
\Gamma[\bar\phi]=\min_J\tilde\Gamma[J,\bar\phi].
\end{equation}
From this one can infer that $\Gamma[\bar\phi]$ is itself a convex functional of $\bar\phi$ \cite{RevModPhys.47.165}. For a constant field $\bar\phi$, the effective potential is defined as
\begin{align}
\label{eq:Veff}
 \Gamma[\bar\phi]=-\int d^4x V_{\rm eff}[\phi],
\end{align}
which implies that $V_{\rm eff}$ is a concave functional. In summary, we have that $W[J]$ is a concave functional, $\Gamma[\bar\phi]$ is convex, and $V_{\rm eff}(\bar\phi)$ is concave. As noted by Weinberg and Wu in reference \cite{PhysRevD.36.2474}, the latter property is not in contradiction with the existence of false, unstable vacua, which in principle require a potential with alternating positive and negative second derivative.  The reason is that the effective potential evaluated at a field-value $\bar\phi$ captures the minimum amount of work needed to change the \emph{groundstate} of the system in the presence of a current enforcing a groundstate expectation value $\langle\phi\rangle=\bar\phi(J)$ \cite{Haymaker:1983xk}. Equivalently, the effective potential can be understood in terms of the \emph{minimum} energy density of states $|s\rangle$ with $\langle s|\phi|s\rangle=\bar\phi$ \cite{PhysRevD.36.2474}. This does not capture the energy density of unstable vacua, but rather that  of the true vacuum. This should be clear from the construction of the true-vacuum effective action starting from the transition amplitude of the true vacuum onto itself. The crucial difference between the cases of the true and false vacuum can be nicely understood from equation \eqref{eq:Gamma}, showing that the true effective action implies summing over sectors in which the wave-function of the true vacuum has a peak. In contrast, usual perturbative calculations rely on a single path integral, which, in the presence of false vacua, will only capture a partial contribution to the effective action, and thus fail to yield a convex result. A relevant example is the evaluation of the Standard Model's effective potential, which for the central values of the Higgs and top masses measured by experiments exhibits an instability, and turns out to be complex and non-convex (see for example \cite{Degrassi:2012ry}). The connection between the unstable one-loop effective potential effective potential and Weinberg and Wu's local effective action was already pointed out in reference  \cite{Einhorn:2007rv}. 

Coming back to the relation between the effective potential and the energy density of quantum states, we would like to remark that, as noted by Weinberg and Wu, one may identify the usual calculations of the effective potential with a minimization of the energy density of states further constrained to have a small dispersion (and in this sense required to be ``local''). In this way, the minimization selects false vacuum states rather than the true vacuum. In the present discussion this can be immediately understood from the fact that a single path integral with boundary conditions $q^{J,m}_0$ corresponds to wave-functions peaking at $q^{J,m}_0$, which gives a simple functional-integral interpretation of Weinberg and Wu's ``localized'' effective potential. As was said earlier, it has been shown in explicit calculations in a variety of works that summing over different path integrals (or equivalently expanding around different saddle-points) it is possible to obtain a concave effective potential \cite{Fujimoto:1982tc,Bender:1983nc,Cooper:1983cd,Tabata:1985hu,Hindmarsh:1985nc,Alexandre:2012hn,Alexandre:2012ht}. Here we have argued that this sum can be understood as a consequence of a multi-peaked vacuum wave-function. The usual constructions of concave potentials involve only summing over the diagonal $m=n$ contributions
in equation \eqref{eq:Gamma}, while our line of reasoning calls for including additional sectors with mixed boundary conditions, corresponding to tunneling effects between the local vacua. These tunneling effects, however, can be nonperturbatively suppressed with respect to the perturbative contributions of the $m=n$ sectors; the situation would be analogous to the inclusion of instanton corrections in gauge theories, corresponding to tunneling in between topological vacua.

As should be clear from the previous discussions, the real, convex functional $\Gamma$ cannot play a role in the computation of tunneling rates. However, one may construct an alternate effective action from the false vacuum transition amplitude $Z^T_F[J]$, and this new functional,  being neither convex nor real, will turn out to play a crucial role for tunneling rates. Moreover, it will be closely related to the usual perturbative evaluations of the effective action yielding complex, non-convex results, as in the Standard Model. In analogous manner to the definition of $\Gamma$, one can define the false vacuum effective action $\Gamma^T_F$ as the Legendre transformation of the false vacuum functional $W^T_F[J]$. 
Note that such a definition assumes a well-defined relation between a source $J$ and a false vacuum mean field $\bar\phi^T_F$. It has been argued that this can be problematic for a non-convex $W^T_F[J]$ --or rather, when there are multiple vacua-- since $\bar\phi^T_F(J)$ and $W^T_F[J]$ may be multivalued at the classical level \cite{Fujimoto:1982tc}, with the multivaluedness arising from the existence of the different classical vacua in the presence of a source. Classical multivaluedness of $\bar\phi^T_F(J)$ is, however, not a problem at the quantum level, when one sums over field configurations. Even for a non-convex $W^T_F[J]$, this gives a well-defined $\bar\phi^T_F(J)=\delta W^T_F[J]/\delta J$  (see \eqref{eq:W}) as expected from the physical intuition that an external current has a well-defined effect on the system. Indeed, denoting the false vacuum state in the Schr\"oedinger picture as $|F,t\rangle$, $\bar\phi^T_F$ is defined in terms of $J$ as
\begin{align}
\bar \phi^T_F(J)=\frac{\left\langle F,T|\phi|F,-T\right\rangle^J}{\left\langle F,T|F,-T\right\rangle^J}.
\end{align}
At the level of the path integral definition in \eqref{eq:WFav}, $\bar\phi^T_F(J)$ only picks up fluctuations associated with the false vacuum, as is clear from the boundary conditions in the integral. In the case of the true vacuum mean field, it is also well defined at the quantum level, as again $\bar\phi$ is unambiguously defined as an expectation value in the true-vacuum. However, in this case 
one may have to sum over different path integrals which capture the different classical branches of the relation between the mean field and the current, as
in equation \eqref{eq:Wav}.  

Despite the well-defined $\bar\phi^T_F(J)$, the existence of a well-defined inverse function $J(\bar\phi^T_F)$ is however not guaranteed, given the non-monotonic behavior of $\delta W^T_F[J]/\delta J$. In this case $J(\bar\phi^T_F)$ will have different branches corresponding to the different vacua, and it has to be ensured that one picks out the branch corresponding to the false vacuum. Again, this is enforced by appropriate boundary conditions. In this way, 
in the single  integral approximation of equation \eqref{eq:WFapprox}, $\Gamma^T_F$ is implicitly defined by 
\begin{equation}
\label{eq:GammaF}
 \exp i\Gamma^T_F[\bar\phi_F;\xi]=\int_{0}^{0} [d\phi] \,\mu(\phi)\exp i\left[S_g[\bar\phi_F,\phi;\xi]-\Gamma^T_{F,j}[\bar\phi_F;\xi]\phi^j\right].
\end{equation}
The zero boundary conditions can be explained as follows. We are assuming that the source enforces the same false vacuum state at $t=\pm T/2$, so that the mean field will approach the same value $\bar\phi_0$ at these times. In a single-path integral approximation, the  false vacuum wave-function in the presence of a source  is then expected to peak at this mean value,  i.e. $q^J_F=\bar\phi_0$. The boundary conditions of the path integral in equation \eqref{eq:GammaF} are the analogues of those in \eqref{eq:Gamma}, but with the $q^{J,m}_{0}$ reduced to a single  $q^J_F$, giving
$q^J_F-\bar\phi_{0}=0$.

Since $W^T_F[J]$ is non-convex, it follows that the resulting effective potential $V_{F\,{\rm eff}}$ (defined similarly to \eqref{eq:Veff}) will be non-convex, and  is expected to have a local minimum related with the false vacuum. In fact $\Gamma_F$ can be understood as one of the aforementioned Weinberg and Wu's ``local'' functionals, corresponding to choosing a ``wrong'' minimization branch in equation \eqref{eq:Gammamin}. Locality, which in reference \cite{PhysRevD.36.2474} was enforced with a restriction in the dispersion of field operators, follows here from the restriction to false vacuum boundary conditions in the path integral. The ``wrong minimization branch'' alludes to choosing the false vacuum state rather than the true vacuum; in our path integral definition, this is again a consequence of the  boundary conditions. When   performing the Legendre transformation, the former picks out the branch of the multivalued function $J(\bar\phi)$ that corresponds to the false vacuum state.

\section{\label{sec:Nielsen}Nielsen identitites}

Once we have defined the effective action functionals $\Gamma$ and $\Gamma_F$, we may further specify the terms appearing inside the path integrals, and study the gauge dependence, following the treatment of reference \cite{Kobes:1990dc}. In equations \eqref{eq:Gamma},  \eqref{eq:GammaF}, $S_g[\bar\phi,\phi;\xi]$ corresponds 
to the gauge-fixed action, given by the sum of the
usual classical action evaluated on $\bar\phi+\phi$,  plus a gauge-fixing term
$S_{gf}[\bar\phi,\phi;\xi]$, and a ghost term $S_{gh}[\bar\phi,\phi,\bar c,c;\xi]$ depending on additional Grassmannian ghost fields $\bar c$, $c$. These  contributions adopt the form
\begin{equation}\label{eq:Sg}
\begin{aligned}
 S_g[\bar\phi,\phi;\xi]=&S[\bar\phi+\phi]+S_{gf}[\bar\phi,\phi;\xi]+S_{gh}[\bar\phi,\phi;\xi],\\
 S_{gf}=&\int d^Dx\, \frac{1}{2\xi} {\cal F}^a {\cal F}^a,\\
 S_{gh}=&\int d^Dx \,\bar c^a \,{\cal H}^{ab}\,c^b,
 \end{aligned}
\end{equation}
where, using the notation of  \eqref{eq:gaugetr} and  omitting the dependence on the fields,
\begin{equation}
\label{eq:H}
{\cal H}^{ab}=F^a_{,k} D^{bk}.
\end{equation}
In the identities \eqref{eq:Sg} and \eqref{eq:H}, ${\cal F}^a$ is the gauge-fixing function, which for example has the form  ${\cal F}^a=\partial_\mu A^a_\mu$ in Fermi gauges ($A_\mu^a$ being the gauge field) though more generally it
may depend on scalar fields and their expectation values, as
in $R_\xi$ gauges.

The Nielsen identities for the effective actions can be derived by studying how the contributions inside the path integrals are modified under changes of $\xi$.  Assuming for simplicity that ${\cal F}^a$ is independent of the gauge-fixing parameter (the result for the general case will be given later) and considering an infinitesimal change of $\xi$, the only contribution inside the path integral that  is modified is $S_{gf}$,
\begin{align}
\label{eq:xitr}
 \delta_\xi S_{gf}=-\frac{d\xi}{2\xi^2}\int\,d^D x {\cal F}^a {\cal F}^a,
\end{align}
while under an infinitesimal gauge transformation with parameter $d\alpha$ it behaves as
\begin{align}
\label{eq:xig}
 \delta_\alpha S_{gf}=\frac{1}{\xi}\int\,d^D x {\cal F}^a {\cal F}^a_{,j}D^{bj} d\alpha^b=\frac{1}{\xi}\int\,d^D x {\cal F}^a {\cal H}^{ab}d\alpha^b.
\end{align}
As noted in \cite{Kobes:1990dc}, the effect of the transformation in \eqref{eq:xitr} can be compensated by appropriately engineering a gauge transformation as in equation \eqref{eq:xig}. 
This happens for a choice of gauge parameter
\begin{equation}
\label{eq:alpha}
 \begin{aligned}
  d\alpha=\frac{d\xi}{2\xi}{\cal H}^{-1} {\cal F}\equiv \frac{d\xi}{2\xi}{\cal G}{\cal F},
 \end{aligned}
\end{equation}
where we defined ${\cal G}={\cal H}^{-1}$.
This gauge transformation does not necessarily leave the path-integral measure invariant; however, as seen in \cite{Kobes:1990dc}, the corresponding Jacobian exactly cancels the variation of the ghost action $S_{gh}$, as long as
the measure satisfies \eqref{eq:mu}.
This may be most easily checked by writing the exponential of the ghost integral as a determinant,
\begin{equation}
 \int [d\bar c][dc]\mu(\bar c,c)\exp iS_{gh}=\det {\cal H}=\exp{\rm tr}\log {\cal H},
\end{equation}
where the trace affects the discrete and spacetime indices of the operator $\cal H$ of equation \eqref{eq:H}. Under the gauge transformation with parameter \eqref{eq:alpha}, the classical action
$S[\phi]$ remains invariant.\footnote{In case of a finite value of $T$, this requires identical boundary conditions for the fields at $t=\pm T/2$ in order to allow integration by parts.} Therefore,  after combining a variation of $\xi$ with a field redefinition given by the gauge transformation of \eqref{eq:alpha}, the net effect in equations \eqref{eq:Gamma}, \eqref{eq:GammaF} 
 is simply a change in the source term, e.g.
$-\Gamma_{,j}[\bar\phi;\xi]\phi^j$. Focusing on $\Gamma$, we may add a further transformation $\phi\rightarrow\phi-\bar\phi$, so as to be able to identify operator averages as in \eqref{eq:av}. A subtlety is that the boundary conditions in the path integrals could themselves depend on the gauge parameter, and they are affected by the gauge transformation \eqref{eq:alpha}. However, it can be easily seen that  the path integrals in the $m=n$ sectors are stationary with respect to infinitesimal variations of the boundary conditions, given that the latter are identical for $t=\pm T/2$. The variations of the $m\neq n$ sectors cancel in pairs if ${\cal N}_{mn}={\cal N}_{nm}$. The ${\cal N}_{mn}$ are related to areas of the vacuum wave-function under its peaks, and thus should be gauge-independent given their probabilistic interpretation; for simplicity we will omit in the following their dependence on the current $\Gamma_{,j}$. Going beyond the discrete sum approximation of equation \eqref{eq:Wsum}, the variations in the boundary conditions in the $\phi$ path integrals under the gauge transformation of equation \eqref{eq:alpha} can be absorbed by redefining the $q,q'$ variables, which should not affect the value of the vacuum wave-function, which should be gauge-invariant. With the previous discussion in mind, we can just ignore the effect of infinitesimal changes in the boundary conditions and write, in the discrete sum approximation (although a similar result will hold for the full effective action)
\begin{equation}
\begin{aligned}
 &\exp i\Gamma[\bar\phi;\xi+d\xi]=\\
 &\sum_{m,n}{\cal N}_{mn}\int_{q^{J,m}_0}^{q^{J,n}_0} [d\phi] \mu(\phi)\exp i\left[\tilde S_g[\phi;\xi]-\Gamma_{,j}[\bar\phi;\xi+d\xi](\phi^j-\bar\phi^j)-\frac{d\xi}{2\xi}\Gamma_{,j}[\bar\phi;\xi]\tilde D^{aj} {\tilde {\cal G}}^{ab}{\tilde{\cal  F}}^b\right],
\end{aligned}
\end{equation}
where we have ignored contributions of higher order in $d\xi$ where appropriate, and $\tilde D^a_j$, $\tilde{\cal G}^{ab}$ and $\tilde{\cal F}^b$ are obtained from $D^a_j$, ${\cal G}^{ab}$ and ${\cal F}^b$ after substituting $\phi\rightarrow\bar\phi-\phi$.
For an infinitesimal $d\xi$  this implies
\begin{align}
\nn \frac{\partial}{\partial\xi}\Gamma=&-e^{-i\Gamma}\sum_{m,n}{\cal N}_{mn} \int_{q^{J,m}_0}^{q^{J,n}_0} [d\phi]\mu(\phi)\left(\frac{\partial}{\partial\xi}\Gamma_{,j}(\phi^j-\bar\phi^j)+\frac{1}{2\xi}\Gamma_{,j}{\tilde D}^{aj} {\cal{\tilde G}}^{ab}{\cal{\tilde F}}^b\right)
 \exp i\left[ \tilde S_g-\Gamma_{,j}(\phi^j-\bar\phi^j)\right]\\
 &=-\left\langle\frac{\partial}{\partial\xi}\Gamma_{,j}(\phi^j-\bar\phi^j)+\frac{1}{2\xi}\Gamma_{,j}{\tilde D}^{aj} {\tilde{\cal G}}^{ab}{\tilde{\cal F}}^b\right\rangle,
\end{align}
where we used the definition of average of equation \eqref{eq:av}.
Using that for the mean field one has $\langle \bar\phi-\phi\rangle=0$ (see \eqref{eq:av} and \eqref{eq:Wav}), 
then the 
effective action  satisfies
\begin{align}
\label{eq:Nielsen}
 \xi\frac{\partial\Gamma}{\partial\xi}[\bar\phi;\xi]+\Gamma_{,j}[\bar\phi;\xi] K^j[\bar\phi,\xi]=0,
\end{align}
with
\begin{align}
\label{eq:K}
 K_j[\bar\phi;\xi]=\left\langle\frac{1}{2}{\tilde D}^a_j {\tilde{\cal G}}^{ab} {\tilde{\cal F}}^b\right\rangle.
\end{align}
Equation \eqref{eq:Nielsen} is the well-known Nielsen identity of the effective action\footnote{The expression for the Nielsen identities in \cite{Kobes:1990dc} involves $D^a_j$, ${{\cal G}}^{ab}$ and  ${\cal F}^b$,  rather
than their counterparts with tildes. This is because we defined the gauge-fixing function within the path integral in \eqref{eq:Gamma}, while Kobes et al's starting point in reference \cite{Kobes:1990dc}  is obtained from \eqref{eq:Gamma} after the field redefinition $\phi\rightarrow\phi-\bar\phi$. Our choice allows to make a more direct contact with the path integral defining the tunneling rate, and its gauge-fixing.}, expressing the fact that the gauge dependence amounts to a nonlocal field redefinition given by $K$ in \eqref{eq:K}. Although in our derivation we assumed that  ${\cal F}^a$  did not depend on $\xi$, it can be seen that if this assumption is relaxed, the formula for $K_j$ becomes
\begin{align}
 K_j[\bar\phi;\xi]=\left\langle\frac{1}{2}{\tilde D}^a_j {\tilde{\cal G}}^{ab} {\tilde{\cal F}}^b-\xi {\tilde D}^a_j{\tilde{\cal G}}^{ab} \frac{\partial \tilde{\cal  F}^b}{\partial\xi}\right\rangle.
\end{align}
As stressed in \cite{Kobes:1990dc}, the gauge-fixing function was kept arbitrary throughout the derivation, the only requirement being that the Faddeev-Popov matrix ${\cal H}$ of \eqref{eq:H} has a well-defined
inverse $\cal G$.

An immediate consequence of the Nielsen identity is that the value of the effective action on the solutions to the equations of motion, 
\begin{align}
\label{eq:eom}
\Gamma_{,i}[\bar\phi;\xi\,]=0,
\end{align}
is gauge-independent. We may further use  \eqref{eq:Nielsen} to understand how the  solutions to \eqref{eq:eom} are affected by a change of the gauge parameter \cite{Nielsen:1975fs}. Let's assume that $\varphi(\xi)$ solves \eqref{eq:eom} for a given $\xi$. Then, taking a functional derivative with respect to $\bar\phi_i$ in \eqref{eq:Nielsen} 
and imposing \eqref{eq:eom} one gets
\begin{align}
\label{eq:der1}
 \left.\left(\xi\frac{\partial}{\partial\xi}+K_j\frac{\delta}{\delta\bar\phi_j}\right)\Gamma_{,i}[\bar\phi;\xi]\right|_{\bar\phi=\varphi(\xi)}=0.
\end{align}
On the other hand, if  $\varphi(\xi)$ solves \eqref{eq:eom} for \emph{all} $\xi$, one should have
\begin{align}
\label{eq:der2}
 \xi\frac{d}{d\xi}\Gamma_{,i}[\varphi(\xi);\xi]=\left.\left(\xi\frac{\partial}{\partial\xi}+\xi\frac{d\varphi_j(\xi)}{d\xi}\frac{\delta}{\delta\bar\phi_j}\right)\Gamma_{,i}[\bar\phi;\xi]\right|_{\bar\phi=\varphi(\xi)}=0.
\end{align}
Comparing equations \eqref{eq:der1} and \eqref{eq:der2} allows to conclude that the solutions to the quantum equations of motion lie along the characteristic curve
\begin{align}
\label{eq:curve}
 \xi\frac{d\varphi_i(\xi)}{d\xi}=K_i[\varphi(\xi);\xi].
\end{align}
 The previous results imply that the vacuum-to-vacuum amplitude in the absence of sources is also gauge independent, since $Z[0]=\exp iW[0]=\exp i\Gamma[\varphi(\xi)]$, where $\varphi(\xi)$ is an extremum satisfying  $J_j=-\Gamma_{,j}[\varphi(\xi)]=0$.

The previous derivation of the Nielsen identities can be repeated for the false vacuum effective action functional $\Gamma^T_F$, with the result
\begin{align}
\label{eq:NielsenF}
 \xi\frac{\partial\Gamma^T_F}{\partial\xi}[\bar\phi;\xi]+\Gamma^T_{F,j}[\bar\phi;\xi] K^j_F[\bar\phi,\xi]=0,\quad K_{Fj}[\bar\phi;\xi]=\left\langle\frac{1}{2}{\tilde D}^a_j {\tilde{\cal G}}^{ab} {\tilde{\cal F}}^b-\xi {\tilde D}^a_j{\tilde{\cal G}}^{ab} \frac{\partial \tilde{\cal  F}^b}{\partial\xi}\right\rangle_F,
\end{align}
where in this case the false vacuum average can be written as
\begin{align}
\label{eq:Fav}
\langle{\cal O}\rangle_F= e^{-i\Gamma^T_F}\int_{q_F}^{q_F}\,[d\phi]\mu(\phi){\cal O}
 \exp i\left[ \tilde S_g[\phi;\xi]-\Gamma^T_{F,j}(\phi^j-\bar\phi^j)\right].
\end{align}
Once more, it follows that the false vacuum effective action is gauge-independent at its extrema, and the analogue of equation \eqref{eq:curve} holds for the extremal  configurations.

\section{\label{sec:tunneling}Tunneling rates}

From the definition of $\Gamma^T_F$ as the Legendre transform of $W^T_F$ one may write the false vacuum transition amplitude in terms of an extremal value of $\Gamma^T_F$, 
\begin{align}
\label{eq:WGammaF}
 Z^T_F[0]=\exp i\Gamma^T_F[\varphi_F(\xi)], \quad {\text{with}}\quad J_j=-\Gamma^T_{F,j}[\varphi_F(\xi)]=0.
\end{align}
This implies that $Z^T_F[0]$ is gauge-independent. From this it automatically follows that the decay rate is gauge-independent as well, as a consequence of equation \eqref{eq:tauG}. The formula for the decay rate can be rewritten as
\begin{align}
\label{eq:tauGf}
\gamma=\lim_{V,T\rightarrow\infty}\frac{2}{VT}\,{\rm Im}\,\Gamma^T_F[\varphi_F(\xi)].
\end{align}
Note that the tunneling rate is associated to an imaginary part in the false vacuum effective action, which, in contrast to the true-vacuum effective action, is complex rather than real.

The result of equation \eqref{eq:WGammaF} can also be obtained by following a derivation closer to \cite{Callan:1977pt}, paying particular attention to boundary conditions. The resulting expressions will be valid whenever equation \eqref{eq:WFapprox} holds. Let's assume that 
a local non-convex effective action $\Gamma^T_F$ has been constructed, whose effective potential shows the appearance of a false vacuum configuration 
$\phi=q_F$. The latter corresponds to a local minimum of the effective potential $V_{F\,{\rm eff}}$, satisfying
\begin{align}
\label{eq:Vmin}
 \left.\frac{\partial V_{\text{F\,eff}}(\phi;\xi)}{\partial\phi_i}\right|_{\phi=q_F}=0.
\end{align}
Note that, given the gauge dependence of $V_{\text{eff}}$, encoded by the Nielsen identities, $q_F$ is itself gauge dependent. Callan and Coleman write the vacuum-to-vacuum transition amplitude as
\begin{align}
 \label{eq:G0}
Z^T_F[0]=\langle q_F|e^{-i HT}|q_F\rangle=\int_{q_F}^{q_F}[d\phi]\mu(\phi)\exp{i\tilde S_{g}[\phi;\xi]},
\end{align}
where, as we discussed at the end of section \ref{sec:vacfun}, it is implicitly  assumed that the wave-function of the false state $|F\rangle$ overlaps maximally with the configuration $q_F$, as we have assumed earlier, and as follows from comparing \eqref{eq:G0} with \eqref{eq:WF0}. To simplify the treatment of the gauge-dependent boundary conditions, 
we may rewrite the fields as 
\begin{align}
\label{eq:fieldredef}
\phi=\varphi_F(\xi)+\rho,
\end{align}
where $\varphi_F(\xi)$ is a fixed configuration satisfying equation \eqref{eq:eom}, with boundary conditions 
\begin{align}
\label{eq:bc}
 \lim_{t\rightarrow\pm T/2}\varphi_F=q_{F},
\end{align}
while the field $\rho$ goes to zero at $t=\pm T/2$.

Then the path integral $W^T_F[0]$ can be rewritten, adding a zero contribution depending on $\Gamma^T_{F,j}[\varphi_F;\xi]\rho^j=0$,\footnote{Recall that $\varphi_F(\xi)$ is chosen to
satisfy the quantum equations of motion $\Gamma^T_{F,i}[\varphi_F(\xi);\xi]=0$.} 
\begin{align}
\label{eq:Gnok}
 Z^T_F[0]=\int_0^0[d\rho]\mu(\rho)\exp{i\left[\tilde S_g[\varphi_F(\xi)+\rho;\xi]-\Gamma^T_{F,j}[\varphi_F(\xi);\xi]\rho^j\right]}.
\end{align}
As said before, $\tilde S_g[\phi;\xi]$ includes the classical action $S[\phi]$, and thus the argument of the exponential in equation \eqref{eq:Gnok} involves $S[\varphi_F(\xi)+\rho]$, exactly as in the case of the path integral that defines the effective action evaluated at $\varphi_F(\xi)$ (see \eqref{eq:GammaF} 
and \eqref{eq:Sg}). Furthermore, the zero boundary conditions in the integral of \eqref{eq:Gnok} match those of the definition of $\Gamma^T_F$ in \eqref{eq:GammaF}. Thus, we recover the relation between $Z^T_F[0]$ and $\Gamma^T_F$ of equation \eqref{eq:WGammaF}, identifying the extremal configuration as one that satisfies \eqref{eq:bc}. Note that consistency with equation \eqref{eq:WGammaF} demands the gauge-fixing in the path integral in \eqref{eq:Gnok} to be the same as the one used to calculate the effective action $\Gamma^T_F$ in \eqref{eq:GammaF} and determine the boundary condition $q_F$ by means of equation \eqref{eq:Vmin}. This also follows from the fact that using the gauge-dependent $\varphi_F(\xi)$ in the path integral \eqref{eq:Gnok}  is implicitly assuming that the gauge-fixing 
enforces the fields to belong to the same slice in the space of orbits of gauge transformations that was chosen for the effective action. As a consequence of this, in the path integral in equation \eqref{eq:G0}, prior to the field redefinition in \eqref{eq:fieldredef}, the correct choice of gauge-fixing function will seem 
unconventional, obtained from the one used in the effective action $\Gamma_F$ in equation \eqref{eq:GammaF} by setting $\bar\phi\rightarrow\varphi(\xi)$ and substituting $\phi\rightarrow\phi-\varphi(\xi)$. The need for such a particular gauge-fixing is possibly the reason that the issue of the gauge dependence of tunneling rates has remained obscure for some time.  The choices of gauge-fixing become  more transparent
when comparing path integrals in terms of fields satisfying identical boundary conditions ($\phi\rightarrow0$ at $t\rightarrow\pm\infty$) as in equations \eqref{eq:GammaF} and \eqref{eq:Gnok}. 
If the  path integral of equation \eqref{eq:Gnok} were to involve a gauge-fixing function $\hat{\cal F}$ different than the function ${\cal F}$ used in the calculation of $\Gamma_F$ and its extremal configuration $\varphi(\xi)$, gauge independence would be lost. Dropping the $T's$ to unclutter the notation and denoting quantities evaluated in different gauges with a superscript ${\cal F}$ or ${\hat{\cal F}}$, in this case we would have that $\varphi^{\cal F}(\xi)$ would not be an extremum of $\Gamma^{\hat{\cal F}}_F$. Using the Nielsen identity \eqref{eq:NielsenF}, it follows that the false vacuum transition amplitude $Z_F^{\hat{\cal F}\cal F}[0]\equiv\exp i\Gamma^{\hat{\cal F}}_F[\varphi^{\cal F}]$ would satisfy
\begin{align}
\label{eq:xidep2}
\xi \frac{d}{d\xi}Z_F^{\hat{\cal F}\cal F}[0]=iZ_F^{\hat{\cal F}\cal F}[0]\left(\xi\frac{\partial}{\partial\xi}\Gamma^{\hat{\cal F}}_F+\xi\,\Gamma^{\hat{\cal F},j}_F\frac{d\varphi^{\cal F}_j(\xi)}{d\xi}\right)=iZ_F^{\hat{\cal F}\cal F}[0]\,\Gamma^{\hat{\cal F},j}_F\left(K^{\cal F}_j-K^{\hat{\cal F}}_j\right),
\end{align}
which is nonzero unless $\hat{\cal F}=\cal F$.

The reader may have noted that, while the usual perturbative calculations of tunneling rates involve an exponential of the classical bounce action, such contribution cannot be readily identified in equation \eqref{eq:tauGf}. The underlying reason is that the extremum of $\Gamma_F$ is not unique, and a sum over extremal configurations is needed. The origin of this exponential can  be made more transparent by modifying the derivation following equation \eqref{eq:G0}. First, let's consider $T\rightarrow\infty$, as is appropriate for computing the tunneling rate with the formula \eqref{eq:tauGf}. Then one may notice that the fields satisfying the boundary conditions in the path integral in \eqref{eq:G0}
belong to different  classes, labelled by the number of times they ``bounce''  from $q_F$ to itself between $t=-\infty$ and $t=\infty$.  Dropping $T$ out of the notation, we might then express $Z_F[0]\equiv Z^\infty_F[0]$ in \eqref{eq:G0} as a sum of path integrals $Z^{(k)}_F$ over the different  sectors, with boundary conditions $q^{k}_F$ for each number $k$ of bounces. The Legendre transform of each $Z^{(k)}_F$, associated with time-dependent sources $J^{(k)}$ which give rise to expectation values of the fields inside the $k$-th class, will define a functional $\Gamma^{(k)}_F$ of the form
\begin{align}
 \label{eq:Gammak}\exp i\Gamma^{(k)}_F[\bar\phi]=&\int_{0}^{0} [d\phi]^k \,\mu(\phi)\exp i\left[S_g[\bar\phi,\phi;\xi]-\Gamma^{(k)}_{F,j}[\bar\phi;\xi]\phi^j\right],
\end{align}
where $[d\phi]^k$ denotes that the fields $\phi+\bar\phi$ are restricted to the $k$-th class.\footnote{Meaning that $\phi$ bounces back and forth from zero $k$ times.} Each  $\Gamma^{(k)}_F$  satisfies a Nielsen-identity analogous to \eqref{eq:NielsenF}, with the averages defined by path integrals within the $k$-th class.\footnote{As the boundary conditions did not play a role in the derivation for the Nielsen identities for $\Gamma$ and $\,\Gamma_F$, one can follow the same reasoning to get identities for $\Gamma^{(k)}_F$.} Within each sector one may define extremal configurations $\varphi^k(\xi)$ satisfying $\Gamma^{(k)}_{F,i}[\varphi^k(\xi)]=0$, and such that $\Gamma^{(k)}_F[\varphi^k(\xi)]$ is gauge-independent as a consequence of the corresponding Nielsen identity. In analogy with equation \eqref{eq:fieldredef}, when expressing $Z_F[0]$ as a sum of path integrals, we may rewrite the fields  inside each sector as $\phi=\varphi^k(\xi)+\rho$, where the extremal configuration $\varphi^k$ satisfies the boundary conditions \eqref{eq:bc}, and with the additional constraint that $\varphi^k$ bounces $k$ times from the vacuum to itself.  Then we may write
\begin{align}
\label{eq:G}
 Z_F[0]=\sum_k\int_0^0[d\rho]^k\mu(\rho)\exp{i\left[\tilde S_g[\varphi^k(\xi)+\rho;\xi]-\Gamma^{(k)}_{F,j}[\varphi^k(\xi);\xi]\rho^j\right]}=\sum_k \exp i \Gamma_F^{(k)}[\varphi^k(\xi)],
\end{align}
where the last identity follows from equation \eqref{eq:Gammak} and the previous argument establishing that the gauge-fixing in $\tilde S_g[\varphi^k(\xi)+\rho;\xi]$ has to be the same as in $S_g[\varphi^k(\xi),\phi;\xi]$. The $\Gamma_F^{(k)}$ are related to the $Z_F^{(k)}$ as follows,
\begin{align}
\label{eq:ZFGamma}
 Z_F^{(k)}[0]= \exp i \Gamma_F^{(k)}[\varphi^k(\xi)].
\end{align}

In order to perform the  sum in $k$ in \eqref{eq:G}, we can resort to the same arguments that were used in \cite{Callan:1977pt}  to show that the contributions of the ordinary semiclassical $k$-bounce solutions exponentiate. The quantum equations of motion are invariant under time translations, and if the bounces are infinitely separated, then the boundary conditions for a $k$-bounce are also invariant under a finite shift in the time coordinate. Thus, time translations of the individual bounces within a $k$-bounce also solve the quantum equations of motion. This implies that fluctuations of $\rho$ given by arbitrary time translations of $\varphi^k$ in \eqref{eq:G} have identical contributions to the path integral. We can define the functional integration on each sector $k$ as an integration over $k$ time translations accompanied by a product of $k$ functional integrations of field excitations with time coordinates restricted to lie around the timestamps $t_1<t_2<\cdots t_{k-1}$ of the centers of the  bounces.  From the previous arguments it follows that the integration over time translations for each bounce simply leads to an overall constant,
\begin{align}
\label{eq:intT}
\int_{-T/2}^{T/2}d t_1\int_{t_1}^{T/2}dt_2\cdots\int_{t_{k-1}}^{T/2}d t_k=\frac{T^k}{k!}.
\end{align}
The remaining functional integrations over the time-constrained field excitations factorize. This is because for a fixed choice of the $t_k$, the time integration in $\tilde S_g$ can be written as a sum of integrals that only depend on the field excitations within each time interval. The factorized contributions correspond to path integrals of fluctuations (excluding time translations) around a single bounce. 
Now, since in equation \eqref{eq:intT} we are integrating over the location of the timestamps, it follows that for a given factorization the time intervals centered around the $t_k$ are not necessarily identical, so that the 
factorized path integrals are in principle different. However, all the bounces approach the constant field configuration $q_F$ at the endpoints of the time intervals, so that the ambiguities coming from the lengths of the time intervals will disappear if
the fluctuations around these endpoints do not contribute. This is guaranteed if the false vacuum effective potential satisfies $V_{F\,{\rm eff}}[q_F]=0$. Indeed, 
 the contributions of the field fluctuations over a time stretch $\hat T$ in which $\varphi^k(\xi)(t)=q_F$ have the form of a vacuum transition functional $Z^{\hat T}_F[q_F]$ defined over the interval ${\hat T}$. $Z^{\hat T}_F$ will have an associated effective action, which will approach $\Gamma_F$ for $\hat T\rightarrow\infty$, and of which the constant configuration $q_F$ is an extremum. Using the same arguments that led to equation \eqref{eq:WGammaF}, and recalling that the effective potential is the zero-momentum piece of the effective action, one would conclude that the contributions of fluctuations around the constant field configurations at the endpoints in between bounces are given by factors of the form $Z^{\hat T}_F[q_F]\sim\exp{i\Gamma_F[q_F]}=\exp(-iV\hat T \,V_{F\,{\rm eff}}[q_F])=1$, where $V$ represents spatial volume, and we used the normalization $V_{F\,{\rm eff}}[q_F]=0$. The former discussion implies that the factorized path integrals around the bounces will be identical, independently of possible ambiguities in the lengths of the time intervals, so that 
\begin{align}                      
\label{eq:sumGk}
Z_F^{(k)}[0]=\frac{T^k}{k!}(\tilde Z_F[0])^k\,\,\Rightarrow \,\,\sum_k Z_F^{(k)}[0]=e^{Z^{(1)}_F[0]}=\exp\exp i\Gamma^{(1)}_F[\varphi^1(\xi)],                                                                                                                                                                                                                                                                                                                                                                                       \end{align}
where we used equation \eqref{eq:ZFGamma}, and we distinguished the contribution of fluctuations exluding time-translations, denoted as $\tilde Z_F^{(k)}[0]$, from the total $Z_F^{(k)}[0]$.
Finally, putting together \eqref{eq:tauGf} and \eqref{eq:sumGk}, we arrive to
\begin{align}
\label{eq:tauM}
 \gamma=-\lim_{V,T\rightarrow\infty}\frac{2}{VT}\,{\rm Im}\,i\, e^{i\Gamma^{(1)}_F[\varphi^1(\xi);\xi]}.
\end{align}
We insist that the former is valid under the normalization $V_{F\,{\rm eff}}[q_F]=0$, which generalizes Callan and Coleman's requirement of a zero classical energy for the false vacuum.
We can obtain a more familiar-looking expression, and make contact with the original results of references  \cite{Coleman:1977py,Callan:1977pt}, by performing an analytic continuation to Euclidean space. Assuming that the  analytic continuation from the real time axis to  $e^{-i\delta}t$ with  $\delta>0$ is unobstructed by any singularities, one may rotate the integral in $S_g$ to the imaginary axis, and formulate the path integrals in 
terms of a gauge-fixed Euclidean action,
\begin{align}
\label{eq:SE}
e^{i\Gamma_F^{(1)}[\varphi^1(\xi)]}\equiv e^{-\Gamma_F^{(1)E}[\varphi^{1,E}(\xi)]}= \int_{0}^0[d\phi]^1\mu(\phi)\exp\left( -S^E_g[\varphi^{1,E}(\xi),\phi;\xi]\right).
\end{align}
$S^E_g[\varphi^{1,E},\phi;\xi]$ is obtained from $-S_g[\varphi^1,\phi;\xi]$ by substituting $t\rightarrow -i\tau$ inside the integrals, and substituting integration in $t$ by integration in $\tau$. In particular,  
the Euclidean configuration $\varphi^{1,E}(\tau,\vec{x};\xi)$ is simply given by the analytic
continuation of $\varphi^{1}$ to imaginary time, i.e. $\varphi^{1,E}(\tau,\vec{x};\xi)=\varphi^1(i t,\vec{x};\xi)$. It thus follows that $\varphi^{1,E}$ satisfies the quantum equations of motion of the Euclidean version of the 
effective action, $\Gamma^{(1),E}$, which can be obtained
from $-\Gamma^{(1)}$ doing the same substitutions that allow to get $S^E$ from $-S$. In terms of the Euclidean effective action and an Euclidean time interval $T^E=i T$, equation \eqref{eq:tauM} becomes
\begin{align}
\label{eq:tauE}
 \gamma=\lim_{V,T_E\rightarrow\infty}\frac{2}{VT^E}\,{\rm Im}\,\, e^{-\Gamma^{(1)E}_F[\varphi^{1,E}(\xi);\xi]}.
\end{align}
An essentially identical formula was obtained in reference \cite{Garbrecht:2015cla}, arising form  a saddle point evaluation of the path integral around a quantum path in theories without gauge fields. In our formalism, we made us of no  saddle point expansion, but rather showed that the use of the quantum path enforces the appropriate boundary conditions in the path integral. Our result was derived  from 
first principles, and on the way we clarified that the effective action involved is a non-convex functional associated with the false vacuum, (rather than the usual effective action, associated with the true vacuum), and accounting for field fluctuations which only bounce once from the false vacuum onto itself. We also established the gauge independence of the result, and  clarified on the way the subtleties related with the compatibility between boundary conditions and gauge-fixing. 

An advantage of our exact results \eqref{eq:tauM}, \eqref{eq:tauE} is that they also clarify how quantum
corrections should be incorporated,  particularly in situations when the tree-level potential has no minima and the saddle point approximation becomes problematic. The path integral has to be indeed evaluated  around a background which solves the quantum equations of motion, which validates the methods of \cite{Garbrecht:2015cla,Garbrecht:2015oea,Garbrecht:2015yza}. An alternative way to get gauge-invariant results for the tunneling rates is to directly compute the false vacuum effective action, including derivative corrections (for example with the methods of references \cite{Chan:1985ny} and \cite{Cheyette:1985ue}), and then solve the quantum equations of motion. A subtlety here is that the usual diagramatic techniques for computing the effective action assume that there are no zero modes. This
is definitely the case when computing $\Gamma_F[\bar \phi]$ for a constant $\bar\phi$. However, for a nontrivial configuration such as the quantum bounce, we expect nontrivial zero modes associated with space-time translations of the center of the bounce (in the constant $\bar\phi$  case, the translated configurations are trivially equivalent to the background, and so there are no such zero modes). Taking into account these zero modes will give a factor $VT$ times a Jacobian, giving
\begin{align}
\label{eq:Jacobian}
 \gamma=2 {\cal J}\,{\rm Im}\,\, e^{-{\Gamma'}^{(1)E}_F[\varphi^{1,E}(\xi);\xi]},
\end{align}
where $\Gamma'$ designates now the effective action obtained by ignoring the zero modes, which coincides with the usual 1PI diagrammatic expansion. Not only the full $\Gamma^E_F[\varphi^E[\xi]]$ is gauge-independent, but the same should happen with $\Gamma'^E_F[\varphi^E[\xi]]$, since it will have its own Nielsen identities and is also extremized by $\varphi^E[\xi]$. Then ${\cal J}=(VT^E)^{-1} \exp[-\Gamma_F+\Gamma'_F]$ cannot depend on the gauge.
The zero modes 
are related to  derivatives of the bounce solution, modulo gauge transformations or field redefinitions. The bounce solution itself is gauge-dependent, see equations \eqref{eq:curve} and \eqref{eq:K}. However, one can construct gauge-independent zero modes by means of a field redefinition.  If the  path-integral measure remains invariant under bosonic field redefinitions (as in dimensional regularization), one can always use the redefined zero modes to construct a gauge-independent Jacobian.\footnote{Note that the field redefinition does not affect the value of $\Gamma^E_F[\varphi^E[\xi]]$, as the bounce is an extremal configuration. Therefore the integration over the zero modes still gives the space-time volume times a Jacobian.} Explicitly, starting from the bounce  $\varphi(\xi)$ one can construct a gauge-independent redefined field configuration $\hat\varphi(\xi,\varphi(\xi))$ satisfying the equation
\begin{align}
\label{eq:redef}
 \xi\frac{\partial\hat\varphi_i}{\partial\xi}+\frac{\partial\hat\varphi_i}{\partial\varphi_j} K_j=0.
\end{align}
Then one may define gauge-independent zero modes as  $\partial_\mu\hat\varphi$, and the Jacobian can be taken as
\begin{align}
\label{eq:Jacobian2}
 {\cal J}=\frac{1}{{\cal M}^4}\prod_{\mu=1}^4\left[\frac{1}{2\pi} \,\partial_\mu\hat\varphi_i\, \partial^\mu\hat\varphi^i\right]^{1/2},
\end{align}
with no summation on $\mu$. ${\cal M}$ is a physical (and thus gauge-independent) mass scale needed for a proper normalization of the measure; for a single field it is given by ${\cal M}^2=V''_{\rm eff}(q_F)$ \cite{Konoplich:1987yd,Garbrecht:2015yza}. The use of the effective potential evaluated at the false vacuum  ensures that the mass scale is  gauge-independent and thus physical.

The action $\Gamma'_F$ and the Jacobian can be computed in a gradient expansion. We may consider for example the case of a real scalar $\sigma$, setting all other mean fields to zero. Then the derivative expansion  will have the form 
\begin{align}
\label{eq:derexp}
 \Gamma'_F[\sigma;\xi]=\int d^4x\left[\frac{1}{2}Z(\sigma;\xi)\partial_\mu \sigma \partial^\mu \sigma-V_{F_\text{eff}}(\sigma;\xi)+O(\partial^4)\right]\equiv\int d^4x {\cal L}_{F_\text{eff}}.
\end{align}
The field redefinition $K_F$ appearing in the Nielsen identities will similarly have a gradient expansion \cite{Metaxas:1995ab,Garny:2012cg},
\begin{align}
\label{eq:Kexpansion}
 K_F(\sigma;\xi)=C(\sigma;\xi)+D(\sigma;\xi)\partial_\mu \sigma \partial^\mu \sigma-\partial^\mu[\tilde D(\sigma;\xi)\partial_\mu\sigma]+O(\partial^4).
\end{align}
Applying these expansions to the Nielsen identity of equation \eqref{eq:Nielsen} yields the identity \eqref{eq:NielsenV} for the effective potential, while for the field renormalization factors
one gets \cite{Metaxas:1995ab,Garny:2012cg}
\begin{align}
\label{eq:NielsenZ}
 \xi\frac{\partial Z}{\partial\xi}=-C\frac{\partial Z}{\partial\sigma}-2Z\frac{\partial C}{\partial\sigma}+2D\frac{\partial V_{F_\text{eff}}}{\partial\sigma}+2\tilde D\frac{\partial^2 V_{F_\text{eff}}}{\partial\sigma^2}.
\end{align}
The identities \eqref{eq:NielsenV} and \eqref{eq:NielsenZ} have been used to argue for the gauge independence of tunneling rates in \cite{Metaxas:1995ab}. There it was assumed that the exponential contribution in the usual formulae for tunneling or nucleation rates involved the effective action of the bounce, rather than its classical action, and gauge independence was shown to follow to lowest nontrivial order with vanishing $D$ and $\tilde D$. This did not clarify the situation at higher orders, aside from the fact that the gauge dependence of the fluctuation determinants in the traditional formulae for the tunneling rate was not addressed. Here we have shown that the false vacuum effective action $\Gamma_F^{(1)}$ evaluated at the bounce 
configuration gives the full answer for the tunneling rate, with no need of including further fluctuation determinants (aside from the zero mode Jacobian). Moreover the gauge independence of the tunneling rate in the derivative expansion follows trivially from  the fact that equations \eqref{eq:NielsenV} and \eqref{eq:NielsenZ} yield the following Nielsen identity for ${\cal L}_{F_\text{eff}}$,
\begin{align}
\label{eq:Leffgauge}
 \xi\frac{\partial}{\partial\xi}{\cal L}_{F_\text{eff}}=\frac{\partial{\cal L}_{F_\text{eff}}}{\partial \sigma}[C+D (\partial\sigma)^2-\partial^\mu(\tilde D\partial_\mu\sigma)]+O(\partial^4),
\end{align}
which vanishes at a solution to the equations of motion of the effective action, $\frac{\partial{\cal L}_{F_\text{eff}}}{\partial \sigma}=0$. 

Regarding the Jacobian ${\cal J}$ in \eqref{eq:Jacobian}, one may use again the derivative expansion at lowest order (as in equation \eqref{eq:derexp}). For simplicity we can consider the case in which
the wave-function renormalization factor $Z(\sigma;\xi)$ is dominated by its field-independent part, $Z(\sigma;\xi)=Z(\xi)$. In this case one can easily solve for the lowest-order contribution to Nielsen's $K$, since \eqref{eq:NielsenZ} implies
\begin{align}
 \xi\frac{\partial Z}{\partial \xi}=-2 Z\frac{\partial C}{\partial\sigma}\Rightarrow C=-\frac{\sigma}{2 Z}\frac{\xi\partial Z}{\partial\xi}.
\end{align}
For a bounce solution $\sigma=\varsigma(\xi)$, we can then construct its gauge-independent redefinition $\hat\varsigma(\xi;\varsigma)$ by solving \eqref{eq:redef}. A possible solution is simply
\begin{align}
 \hat\varsigma= Z^{1/2}\varsigma.
\end{align}
For a Euclidean bounce with $O(4)$ symmetry, the Jacobian \eqref{eq:Jacobian2} can be written as
\begin{align}
 {\cal J}=\left[\frac{1}{8\pi V''_{\rm eff}(q_F)}\int d^4x\,  Z\partial_\mu \varsigma\,\partial^\mu\varsigma\right]^2,
\end{align}
where this time there is summation in $\mu$.
As was done in \cite{Callan:1977pt}  for the classical Euclidean action, using the fact that the effective action $\Gamma'_F$ does not change under infinitesimal deformations of the bounce solution, 
and considering deformations generated by coordinate dilatations $\varphi(x)\rightarrow \varphi(e^{-a }x)$, one can show that to this level of approximation
\begin{align}
\delta\Gamma'_F=0&\Rightarrow\int d^4x\left[ Z\partial_\mu \varsigma\,\partial^\mu\varsigma-4 V_{F_\text{eff}}(\varsigma)\right]=0,\\
&\Rightarrow \Gamma'_F[\varsigma]=\int d^4x\left[ \frac{1}{2}Z\partial_\mu \varsigma\,\partial^\mu\varsigma- V_{F_\text{eff}}(\varsigma)\right]=\frac{1}{4}\int d^4x 
Z\partial_\mu \varsigma\,\partial^\mu\varsigma,
\end{align}
and thus
\begin{align}
\label{eq:Jexp}
 {\cal J}\sim\left[\frac{{\Gamma'}^{(1)E}_F[\varphi^{1,E}(\xi);\xi]}{2\pi V''_{\rm eff}(q_F)}\right]^2.
\end{align}
As expected, the Jacobian is gauge-independent, and in the semiclassical limit $\Gamma^E_F\sim S^E$ one recovers the usual factor appearing in Callan and Coleman's tunneling formula. Indeed, the latter reads
\begin{align}
\label{eq:CC}
 \gamma\sim\frac{(S^E[\phi_b])^2}{4\pi^2}e^{-S^E[\phi_b]}\left|\frac{\det'[-\partial^2+V''(\phi_b)]}{\det[-\partial^2+V''(q_F)]}\right|^{-1/2},
\end{align}
where $\phi_b$ is the bounce solution obtained from the classical potential, ${\rm det'}$ refers to a determinat with zero modes excluded, and $V''(\phi)$ denotes the second-order coefficient in the expansion of the classical potential around the field configuration $\phi$. As has already been noted before, (see for example \cite{Scrucca,Lalak:2016zlv}), the product of the exponential and determinants in \eqref{eq:CC} can be expressed as ${\cal M}^{-4}\exp(-\Gamma'^E[\phi_b])$, so that one gets a formula analogous to our expression \eqref{eq:Jacobian}, but with the Jacobian substituted in terms of the classical Euclidean action, and with the classical bounce playing the role of the quantum bounce, and an additional factor $1/2$ coming from Callan and Coleman's identification of the imaginary part by analytic continuation of the potential. Our results improve on \eqref{eq:CC} by clarifying that both the classical bounce and the Jacobian have to be generalized in terms of a quantum bounce. These two changes are crucial to ensure gauge-independence of the result. Since the bounce action and Jacobian in \eqref{eq:CC} differ from ours by $O(\hbar)$ effects, we can interpret \eqref{eq:CC} as a semiclassical approximation to the more exact result \eqref{eq:Jacobian}.  Also, in our formalism the identification of the imaginary part in \eqref{eq:Jacobian} does not have to involve an unphysical analytic continuation of the classical potential, but can instead be understood as a consequence of unitarity, which as noted in \cite{PhysRevD.36.2474} enforces imaginary parts in  convex regions of the false-vacuum effective potential. 
This holds because at zero  momentum the effective action develops an imaginary part whenever some particle masses become negative, allowing for spontaneous particle production from the false vacuum, signalling a decay. Negative scalar masses are guaranteed in a given background whenever the potential is convex, as the former can be related to the second derivatives of the potential.

Aside from the previous links with  the formalisms of references \cite{Garbrecht:2015cla} and \cite{PhysRevD.36.2474}, we may  make contact with the tunneling formula recently derived in \cite{Andreassen:2016cff}. In this work, the tunneling rate in quantum field theories is written as
\begin{align}
\label{eq:Schw}
 \gamma=\frac{2}{V}\frac{{\rm Im} \int [d\phi]e^{-S^E}\delta(\tau_\Sigma[\phi])}{\int [d\phi]e^{-S^E}}.
\end{align}
In the previous formula, the $\delta$ function enforces integration over field configurations that reach a given surface $\Sigma$ at a time $\tau$.
It is argued in \cite{Andreassen:2016cff} that the imaginary part only comes from the numerator. In our formalism we would interpret the denominator as $Z^E_F[0]=\exp(-\Gamma^E_F[\varphi])$, for some extremal configuration
$\varphi$. If there is no imaginary part involved, this can only be the constant configuration sitting at the false vacuum $\varphi=q_F$. This can be justified with the unitarity arguments mentioned before, which imply that the effective potential can only get an imaginary part when some masses become negative. This cannot happen on a stabilized (up to tunneling effects) false vacuum. Bounce solutions, on the other hand, traverse regions in which the potential is convex and develops an imaginary part. In our normalization, the denominator in \eqref{eq:Schw}
then becomes $\exp(-\Gamma^E_F[q_F])=\exp(-V\hat T \,V_{F\,{\rm eff}}[q_F])=1$. Regarding the numerator, using time translation invariance one may write
\begin{align}
  \int [d\phi]e^{-S^E}\delta(\tau_\Sigma[\phi])=\frac{1}{T}\int [d\phi] d\tau e^{-S^E}\delta(\tau_\Sigma[\phi])\sim\frac{1}{T}\int [d\phi]  e^{-S^E},
\end{align}
where the last equality comes from the fact that integrating over configurations which reach a given surface at any possible time should be analogous to integrating over all field configurations. Again, one may express the path integral in terms of the effective action evaluated at an extremum, but now one giving a nonzero imaginary part, i.e. a bounce solution. In this manner formula \eqref{eq:tauE} is recovered.


Finally, we note that our results have a straightforward generalization to finite temperature thermal tunneling, since in this case the effective action and the vacuum-to-vacuum amplitude still have a path integral formulation (see for example the review \cite{Quiros:1999jp}). This is similar to the Euclidean formulation at zero temperature, but with the fields having (anti)-periodic boundary conditions in the time direction. All 
the formal manipulations of the path integrals employed to arrive to our results at zero temperature can be reproduced in the finite temperature case. 

Although for simplicity we wrote most of the identities in the discrete-sum-approximations of equations \eqref{eq:Wsum}, \eqref{eq:WFapprox},  the result of equation \eqref{eq:tauGf} linking the decay rate of a false vacuum to its associated effective action is valid beyond this simplification, as it just follows from the definition of the Legendre transformation. As argued before, we expect the same general validity for the Nielsen identities and the results concerning gauge-independence derived from them. The formulae \eqref{eq:tauM}, \eqref{eq:tauE}, in turn, are only valid in the limit in which $Z_F[0]$ can be approximated by a single path integral, as in equation \eqref{eq:WFapprox}. In the most general situation, instead of depending on a $k=1$  bounce solution with simple boundary conditions fixed by the false vacuum, the tunneling rate can be expressed as an integration over extrema of $k=1$ effective-action-like functionals with different boundary conditions, weighed by the false vacuum wave-function $\psi_F$ appearing in \eqref{eq:WF}. More explicitly, using the same reasoning leading to \eqref{eq:tauM}, one has in this case
\begin{align}
 \gamma=-\lim_{V,T\rightarrow\infty}\frac{2}{VT}\,{\rm Re}\log\int [dq][ dq']\mu(q)\mu(q')\psi^J_F(q')\psi^{J\star}_F(q)\exp\exp i \Gamma_F^{(1)q'q}[\varphi^{1,q'q}].
\end{align}
In the previous equation, $ \Gamma_F^{(1)q'q}$ denotes a functional defined from a path integral analogous to equation \eqref{eq:GammaF}, but with the $q_F$ in the boundary conditions replaced by $q',q$, and with the integration restricted to field fluctuations in the $k=1$ class. The configurations $\varphi^{1,q'q}$ are extrema of the $ \Gamma_F^{(1)q'q}$, approaching $q',q$ at negative and positive infinite time, respectively, and bouncing only once in between the boundary values.

\section{\label{sec:conclusions}Conclusions}

In this paper we have clarified issues concerning the gauge independence of tunneling and nucleation rates, as well as the question of how to consistently incorporate quantum corrections in their calculation. We have also shed light on  the role played by effective action functionals, paying attention to their convexity properties. These aspects are relevant for allowing unambiguous physical answers in
the study of questions such as the stability of the Standard Model vacuum, or the properties of phase transitions in the early Universe, which can be important for understanding the mechanisms behind baryogenesis.

For some time it has been generally accepted that somehow the quantum effective potential plays a role in the computation of tunneling probabilities. This idea is problematic for two reasons. First,  the effective potential is known to be gauge-dependent. Although this dependence cancels out in physical quantities defined at the extrema  of the potential, such as vacuum energies and scalar masses, the gauge-dependence could taint the usual computations of tunneling rates, which are sensitive to the values of the potential in between minima. On the other hand, the idea that the effective potential plays a role in quantum tunneling goes against the known fact that the true effective potential of the theory is known to be concave, having thus no false minima.

A clear understanding of how to extract a gauge-independent physical result for tunneling rates was lacking, despite hints in some perturbative calculations. This is also related to the problem of consistently including quantum corrections,  which is best illustrated by scenarios in which it is unavoidable to consider quantum corrections to the potential in order to determine the presence of false vacua. It is not straightforward to include these effects in the calculation of tunneling rates using the usual formalism without incurring in a double counting of quantum corrections. 

It turns out that the former problems, the gauge dependence of tunneling rates and the consistent inclusion of quantum fluctuations, have a remarkable simple solution. Starting from the false vacuum transition amplitude onto itself, $Z_F[0]$,  it can be seen that the decay rate is exactly determined by a false vacuum  effective action functional, $\Gamma_F$, evaluated at a solution to its quantum equations of motion. Gauge-independence is immediate from the fact that the Nielsen identities imply that the value of $\Gamma_F$ at its extrema does not depend on the gauge parameters.  $\Gamma_F$ differs from the true effective action of the theory, $\Gamma$, and is complex and non-convex, so that the associated effective potential $V_{F\,{\rm eff}}$ can have a false vacuum, and $\Gamma_F$ an imaginary part, without running into inconsistencies. In fact, $\Gamma_F$ represents one of the ``localized'' effective actions proposed by Weinberg and Wu \cite{PhysRevD.36.2474}, where in this case the restriction of the field fluctuations is enforced by the localization of the field-space wave-function of the false vacuum state. In regards to the true-vacuum effective action, we have shown that the need to sum over path integrals in convex constructions of $\Gamma$ is due to a multi-peaked wave function of the groundstate.

In Euclidean space, in the approximation in which the false vacuum transition amplitude reduces to a single path integral, this means
\begin{align}
\label{eq:tauconc}
 \gamma=\lim_{V,T^E\rightarrow\infty}\frac{2}{VT^E}\,{\rm Im}\,\, e^{-\Gamma^{(1)E}_F[\varphi^{1,E}(\xi);\xi]}=2 {\cal J}\,{\rm Im}\,\, e^{-{\Gamma'}^{(1)E}_F[\varphi^{1,E}(\xi);\xi]}.
\end{align}
In the equations above,  $T^E$ is the Euclidean time interval, and the false vacuum effective action $\Gamma^{(1)E}_F$ is defined as the Legendre transform of the contribution $Z^{(1)}_F[0]$ to $Z_F[0]$ which arises from field fluctuations involving a single infinite-time bounce from the false vacuum onto itself. The configuration $\varphi^{1,E}(\xi)$ appearing in \eqref{eq:tauconc} is a generalized bounce configuration that solves the equation $\Gamma^{(1)E}_{F,i}=0$. This solution must approach the false vacuum configuration $q_F$ that minimizes the effective potential at Euclidean times $\tau\rightarrow \pm\infty$, 
and the superscript $1$ in $\varphi^{1,E}(\xi)$ reflects the requirement that the field configuration should only bounce once in between the minimum configurations. The effective potential is assumed to be defined in such a way that it vanishes at the false vacuum. On the right-hand-side of \eqref{eq:tauconc}, ${\Gamma'}^{E}_F$ denotes the Euclidean effective action without the integration over zero modes, which coincides with the usual diagrammatic expansion. ${\cal J}$ is a Jacobian that at lowest order in a derivative expansion is given by \eqref{eq:Jexp}. 
The fact that ${\Gamma'}^{E}_F$ includes a non-convex effective potential allows to understand the origin of the imaginary part from the usual unitarity arguments in quantum field-theory. Convex regions of the effective potential imply negative scalar masses, which gives an imaginary part to the effective action even at zero momentum. At one-loop, this arises from the logarithms of the effective masses in the Coleman-Weinberg formula for the effective potential.

The false vacuum effective action evaluated at the bounce configuration already includes all quantum corrections, and aside from the zero-mode Jacobian there is no need to include additional fluctuation determinants. 
From our results it follows that consistent evaluations of tunneling rates can be performed  by computing the false vacuum effective action, including derivative terms (using for example the techniques of \cite{Chan:1985ny,Cheyette:1985ue}), and solving for the quantum bounce. Alternatively, one may use the method of external sources of references \cite{Garbrecht:2015cla,Garbrecht:2015oea} to directly obtain the effective action evaluated at the bounce (see also~\cite{Garbrecht:2015yza}). Since the cancellation of the gauge dependence is automatic, much like in the computation of S-matrix elements, there is in principle no need to perform a field redefinition in the effective action to remove the explicit gauge dependence. Rather, consistent physical results arise after properly accounting for derivative terms in the effective action.\footnote{However, a field redefinition of the gauge-dependent bounce solution might be needed to compute the Jacobian of the zero modes beyond the result of equation \eqref{eq:Jexp}.} In a truncated perturbative expansion, order-by-order gauge independence may require appropriate resummations, as it is known to happen with the energies at the minima of the effective potential \cite{Andreassen:2014eha}, which are formally gauge-independent. It remains to be seen whether order-by-order gauge independence can be achieved for tunneling rates. 

\section*{Acknowledgements}

We would like to thank Nikita Blinov, David Morrissey and  Brian Shuve for useful discussions and feedback. A.P.\ acknowledges support from CONACyT, and C.T.\ from the Spanish Government through grant FPA2011-24568 (MICINN). 

\bibliographystyle{JHEP}  
\bibliography{bibtunneling}  

\end{document}